\documentclass[%
 reprint,
 amsmath,
 amssymb,
 pra,
]{revtex4-2}

\usepackage{dcolumn}
\usepackage{bm}
\usepackage{mathtools}
\usepackage{amsthm} 

\usepackage{graphicx} 
\usepackage{float} 
\usepackage{hyperref}

\usepackage[labelformat=simple]{subcaption} 

\makeatletter
\newcommand{\dalembertian}{\mathop{\mathpalette\dalembertian@\relax}}
\newcommand{\dalembertian@}[2]{%
  \begingroup
  \sbox\z@{$\m@th#1\square$}%
  \dimen0=\fontdimen8
    \ifx#1\displaystyle\textfont\else
    \ifx#1\textstyle\textfont\else
    \ifx#1\scriptstyle\scriptfont\else
    \scriptscriptfont\fi\fi\fi3
  \makebox[\wd\z@]{%
    \hbox to \ht\z@{%
      \vrule width \dimen0
      \kern-\dimen0
      \vbox to \ht\z@{
        \hrule height \dimen0 width \ht\z@
        \vss
        \hrule height 2\dimen0
      }%
      \kern-2.5\dimen0
      \vrule width 2.5\dimen0
    }%
  }%
  \endgroup
}
\makeatother

\begin{document}

\title{Slow and Stopped Light in Dynamic Moir\'e Gratings}
\author{Thomas E.~Maybour}
\author{Devin H.~Smith}
\author{Peter Horak}
\affiliation{Optoelectronics Research Centre, University of Southampton, Southampton~SO17~1BJ, UK}


\begin{abstract}

We investigate a theoretical model for a dynamic Moir\'e grating which is capable of producing slow and stopped light with improved performance when compared with a static Moir\'e grating. A Moir\'e grating superimposes two grating periods which creates a narrow slow light resonance between two band gaps. A Moir\'e grating can be made dynamic by varying its coupling strength in time. By increasing the coupling strength the reduction in group velocity in the slow light resonance can be improved by many orders of magnitude while still maintaining the wide bandwidth of the initial, weak grating. We show that for a pulse propagating through the grating this is a consequence of altering the pulse spectrum and therefore the grating can also perform bandwidth modulation. Finally we present a possible realization of the system via an electro-optic grating by applying a quasi-static electric field to a poled $\chi^{(2)}$ nonlinear medium.      

\end{abstract}

\maketitle


\section{\label{sec:level1}Introduction}

There has been a great amount of interest in slow light since it was demonstrated by Hau et al.\ \cite{Hau1999,Liu2001,Phillips2001} that a laser pulse could be brought to almost complete stand-still by using electromagnetically induced transparency (EIT). It has been a longstanding goal to replicate those results in a solid-state photonic crystal device \cite{Krauss2008}.

All slow light devices work by the same principle: incoming light is coupled to a resonance which impedes the passage of light and results in a reduction in its group velocity \cite{Khurgin2010}. In the case of EIT the light is coupled to atomic electron transitions,  while in a photonic crystal the light is coupled into forward and backward modes via a narrow-band Bragg effect. In either case, there is a trade off between the achievable slow-down and the bandwidth of the resonance $\Delta f$; larger slow-downs correspond to narrower resonances. The coupling strength between light and the resonance also impacts on the performance of the device: A weaker coupling will require a narrower resonance to achieve the same slow-down as a stronger coupling. A common figure of merit to characterize the performance of a slow light device is the delay-bandwidth product (DBP). A slow light device of length $L$ and group velocity $v_{g}$, has an induced delay given by $\Delta t = L / v_{g}$. The DBP is then simply $\Delta t\Delta f$, which is approximately constant for any given slow light device. There have been a number of different papers discussing the DBP limitations of different devices \cite{914401,Tucker2005,KhurginBuffers}. Unfortunately the initial optimism surrounding slow light devices for developing optical delay lines, memories and other devices has ultimately not come to fruition due to the DBP. In practice, an optical fiber offers little slow-down but has a wide acceptance bandwidth and can be made arbitrarily long which gives optical fiber loops a DBP superior to slow light devices. 

In order to overcome the delay-bandwidth limitation time-varying resonant structures have been suggested \cite{Yanik2005-2}. Such structures are based on dynamically varying the bandwidth of the resonance so that initially the structure has a larger bandwidth and higher group velocity, then it is dynamically altered to a small bandwidth and lower group velocity resonance. It has been shown that dynamically varying a slow light resonance in this way with a pulse trapped inside the structure will result in the pulse bandwidth being compressed and the pulse being brought to a practical stand-still \cite{Yanik2004,Yanik2005}, which is analogous to the behavior observed in EIT \cite{Yanik2004-2}. This behavior has been studied in a general theoretical context \cite{Khurgin:05,Daniel2011} and has been applied to various specific devices such as a quasi-phase-matched waveguide using backward frequency conversion \cite{PhysRevA.72.023810}, a p-i-n integrated photonic crystal nanocavity \cite{Tanabe2010}, a waveguide with moving index fronts \cite{Gaafar2020} and in a grating coupled metal-dielectric-metal waveguide \cite{Weiss2020}.

In this work we examine the time-variation of a Moir\'e grating with a slow light transmission band. A Moir\'e grating is created by superimposing two Bragg gratings with different grating periods. Provided the difference in the periods is small, the dual grating structure creates a narrow slow light resonance between two band gaps \cite{MFG}. We show that a laser pulse localized inside a Moir\'e grating can be brought to a standstill by dynamically increasing the grating coupling strength. We find in accordance with previous work \cite{Yanik2004} that dynamic slowing of the pulse corresponds to spectral compression and by applying the time-variation symmetrically and adiabatically the pulse is returned to its previous state. We then show that by applying the time-variation asymmetrically the bandwidth of the pulse exiting the grating can be compressed or broadened. Finally we examine a possible realization by using an electro-optic grating in a quasi-phase matched $\chi^{(2)}$ nonlinear material. 


\section{\label{sec:level2}Background}

\begin{figure*}[tb]
    \centering
    \begin{subfigure}{.33\textwidth}
        \includegraphics[width=\linewidth]{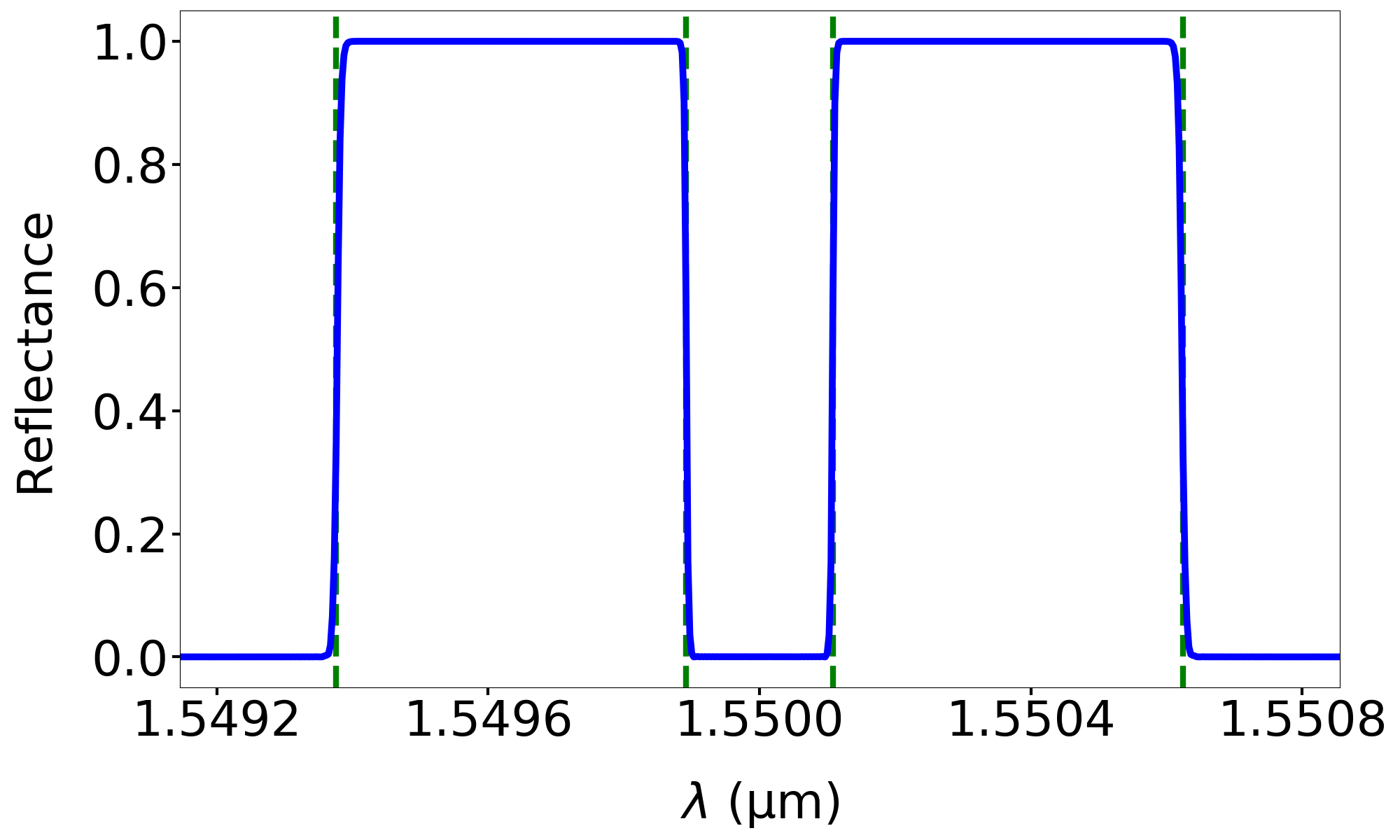}
        \caption{}
        \label{fig:refl}
    \end{subfigure}\hfill
    \begin{subfigure}{.33\textwidth}
        \includegraphics[width=\linewidth]{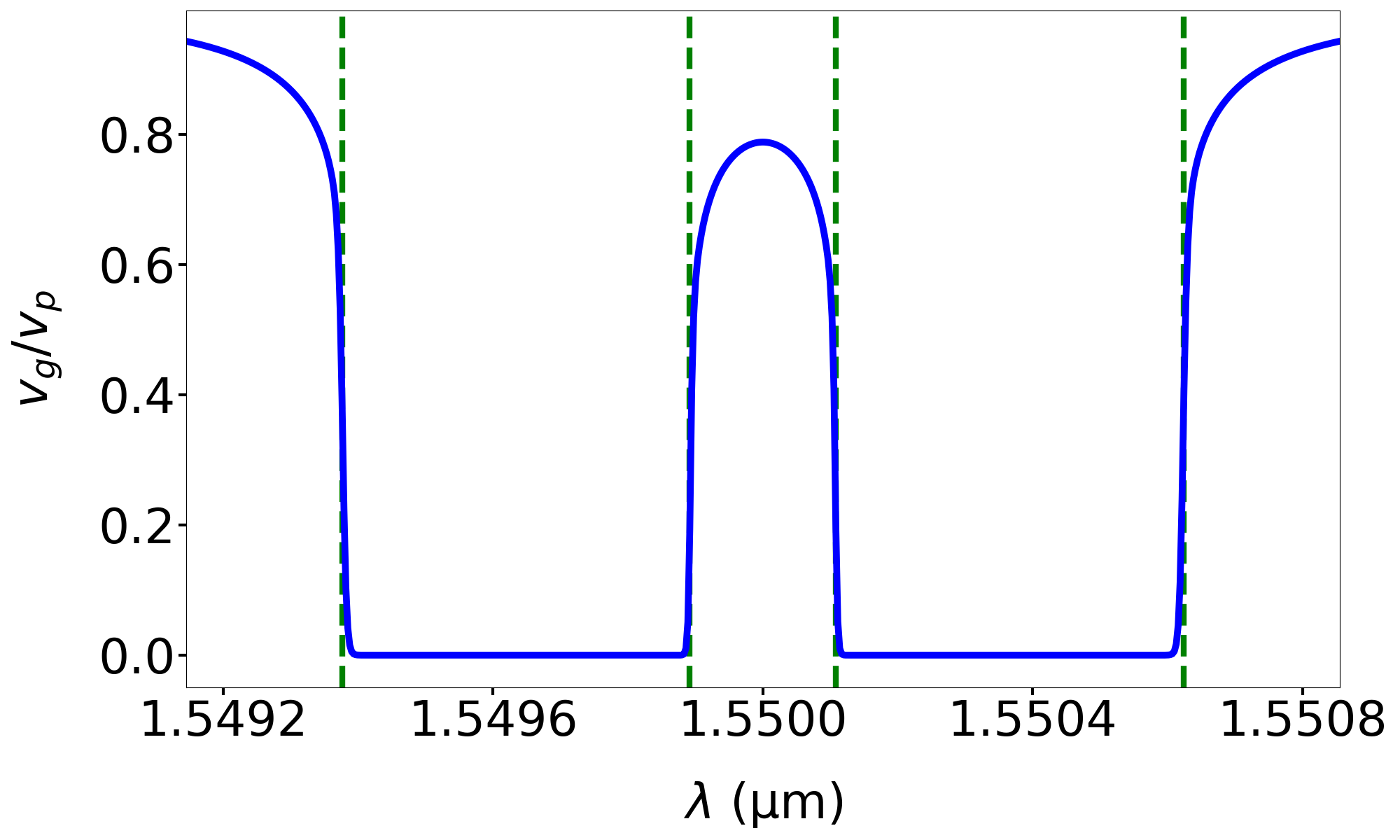}
        \caption{}
        \label{fig:gv}
    \end{subfigure}\hfill
    \begin{subfigure}{.33\textwidth}
        \includegraphics[width=\linewidth]{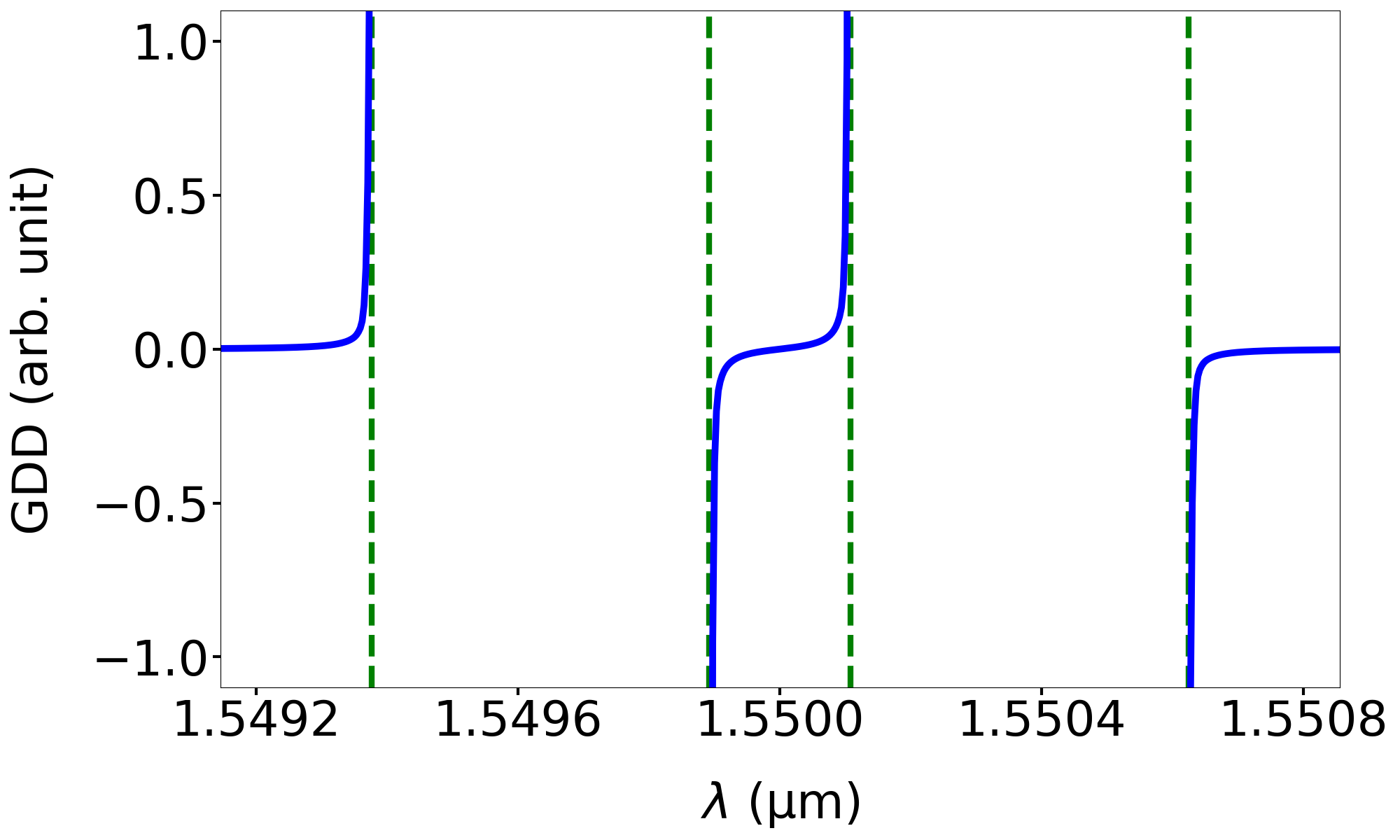}
        \caption{}
        \label{fig:gdd}
    \end{subfigure}
    \caption{\subref{fig:refl} Reflection spectrum, \subref{fig:gv} group velocity and \subref{fig:gdd} group delay dispersion for a Gaussian apodized Moir\'e grating with parameters $\bar{n} = 1.445$, $\delta n = 10^{-3}$, $\lambda_{B} = 1550$\,nm, $\Lambda_{s} = 2.68$\,mm and $L = 10$\,cm.}
    \label{fig:moire_plots}
\end{figure*}

It is a well known feature of Bragg gratings that strong dispersion at the band edges of the Bragg resonance can generate slow light \cite{SLBGA}. Intuitively this is understood by light inside the grating being scattered backwards and forwards before emerging at the far end of the grating, leading to increased propagation time and a reduction in group velocity. Therefore, to achieve a low group velocity the carrier frequency must be tuned close to the band edge in order to increase scattering. The closer it is tuned to the bandgap the lower the achievable group velocity but the pulse spectrum must remain outside the bandgap in order to allow transmission. Consequently lower group velocities have smaller available bandwidth resulting in the familiar trade off between bandwidth and delay. Along with a reduction in group velocity, the band edges generate strong higher order dispersion effects which cause pulse broadening and walkoff. The short wavelength edge of the Bragg resonance generates normal dispersion and the long wavelength edge anomalous dispersion. Moir\'e gratings were suggested \cite{MFG} as a way to exploit this feature of Bragg gratings to cancel second order dispersion and minimize pulse broadening to create a grating better suited to slow light propagation.

A Moir\'e grating is created in the linear refractive index by superimposing two gratings with periods $\Lambda_{1}$ and $\Lambda_{2}$. In one dimension this can be modeled by
\begin{equation}\label{moire_linear_sum}
    n(z) = \bar{n} + \frac{\delta n}{2} \Big(\cos(\alpha_{2}z) - \cos(\alpha_{1}z) \Big)
\end{equation}
where $\bar{n}$ is the AC effective refractive index, $\delta n$ is the grating strength, $\alpha_{1} = 2\pi/\Lambda_{1}$ and $\alpha_{2} = 2\pi/\Lambda_{2}$, and we assume $\Lambda_2>\Lambda_1$. 

The addition of an extra grating period in Eq.~(\ref{moire_linear_sum}) from what would otherwise be a standard Bragg grating generates two Bragg resonances at the wavelengths $\lambda_{1} = 2n_{1}\Lambda_{1}$ and $\lambda_{2} = 2n_{2}\Lambda_{2}$, where $n_{1}$ and $n_{2}$ are the refractive indices at wavelengths $\lambda_{1}$ and $\lambda_{2}$. Figure \ref{fig:refl} shows the reflectance of a Moir\'e grating where the dual bandgaps separated by a transmission band can be clearly seen. Note that Fig.\ \ref{fig:moire_plots} serves as a schematic here; it was calculated using coupled mode theory as discussed later in this section.

The normal and anomalous dispersion generated by the two bandgaps cancel at the center of the transmission band as can be seen in Fig.~\ref{fig:gdd}. This allows pulses to propagate at this wavelength with minimal pulse broadening while still experiencing the reduction in group velocity, as shown in Fig.~\ref{fig:gv}.  

When the Bragg resonances are sufficiently spectrally separated they can be modeled as two independent Bragg gratings. However, this breaks down as the bandwidth separating the resonances decreases sufficiently, at which point they must be modeled as a single structure, as discussed further in section \ref{sec:level3}. One problematic consequence of decreasing the size of the transmission band is that spectral sidelobes generated by the bandgaps increase reflectance within the transmission band. It is therefore necessary to apply an apodization to the grating to suppress the sidelobes and to improve transmission by reducing reflectivity. An apodization can be incorporated into the model by making the grating strength in Eq.~(\ref{moire_linear_sum}) spatially dependent,
\begin{equation}\label{grating_strength}
    \delta n(z) = \delta nf(z)
\end{equation}
where $f(z)$ is a suitable apodization profile. It is convenient to express the Moir\'e grating (\ref{moire_linear_sum}) as the product of two sine waves
\begin{equation}\label{moire_linear}
    n(z) = \bar{n} + \delta n(z) \sin(\alpha_{s}z)\sin(\alpha_{B}z)
\end{equation}
with wavenumbers $\alpha_{B} = 2\pi/\Lambda_{B}$ and $\alpha_{s} = 2\pi/\Lambda_{s}$. The grating periods $\Lambda_{s} = 2\Lambda_{1}\Lambda_{2}/(\Lambda_{2} - \Lambda_{1}) $ and $\Lambda_{B} = 2\Lambda_{1}\Lambda_{2}/(\Lambda_{2} + \Lambda_{1})$ are the Moir\'e and Bragg periods, respectively, with $\Lambda_{S} > \Lambda_{B}$, i.e., the Moir\'e term $\sin(\alpha_{s}z)$ oscillates slower than the Bragg term $\sin(\alpha_{B}z)$. 

In Eq.~(\ref{moire_linear}) the Bragg and Moir\'e terms serve two different physical functions; the Bragg term creates a wide bandgap centered on the wavelength $\lambda_{B} = 2n_{B}\Lambda_{B}$ where $n_{B}$ is the refractive index at $\lambda_{B}$. The Moir\'e term opens a transmission band at $\lambda_{B}$ within the bandgap. The size of this transmission band is dependent both on $\Lambda_{S}$ and $\delta n$ and is generally much narrower than the bandgap of the Bragg term.  

To understand why the Moir\'e term opens a transmission band we re-write Eq.\ (\ref{moire_linear}) as
\begin{equation}\label{moire_linear_phase_shifts}
    n(z) = \bar{n} + \delta n(z)\lvert\sin(\alpha_{s}z)\rvert\sin\Big(\alpha_{B}z + \pi\theta\big(-\sin(\alpha_{s}z)\big)\Big)
\end{equation}
where $\theta$ is the Heaviside step function. The modulus of the Moir\'e term, $\lvert\sin(\alpha_{s}z)\rvert$, thus acts as an envelope to the Bragg grating between $\pi$ phase shifts induced by the term $\theta\big(-\sin(\alpha_{s}z)\big)$. If we consider a grating that has length $L$ and set the Moir\'e period so that $\Lambda_{S} = L$, then the grating will have a single $\pi$ phase shift at its center. If the envelope term $\lvert\sin(\alpha_{s}z)\rvert$ is neglected then the subsequent grating is equivalent to a standard $\pi$ phase shift grating. It is a well known feature of a $\pi$ phase shift grating that a small transmission band opens at the center of the Bragg resonance \cite{Erdogan1997}. As the Moir\'e period is decreased the number of $\pi$ phase shifts increases and the transmission bandwidth correspondingly broadens. 

In addition to the Moir\'e grating, Eqs.\ (\ref{moire_linear})-(\ref{moire_linear_phase_shifts}), we therefore introduce the periodically $\pi$ phase-shifted grating
\begin{equation}\label{moire_pi_shift1_grating}
    n(z) = \bar{n} + \delta n(z) \; \text{sgn}\big(\sin(\alpha_{s}z)\big)\sin(\alpha_{B}z).
\end{equation}
This grating is simply a Bragg grating with $\pi$ phase shifts controlled by the Moir\'e period. Its advantage over the Moir\'e grating is that turning the Moir\'e term $\sin(\alpha_{s}z)$ into a square wave gives a higher average grating strength and correspondingly a higher average coupling strength which improves its slow light performance. The trade off is that without the envelope between the phase shifts, they act as small interfaces which generate reflections.

Both the Moir\'e and $\pi$ phase-shifted  grating are examples of superstructure gratings which can be written in the general form 
\begin{equation}\label{superstructre_grating}
    n(z) = \bar{n} + \delta n(z)a(z)\sin(\alpha_{B}z)
\end{equation}
where $a(z)$ is the superstructure envelope. For the Moir\'e grating the superstructure envelope is $a(z)=\sin(\alpha_{s}z)$ and for the  $\pi$ phase-shifted grating $a(z)=\text{sgn}\big(\sin(\alpha_{s}z)\big)$. 

We model such superstructure gratings using coupled mode theory. This is done by choosing a linearly $x$-polarized electric field ansatz which is composed of forward and backward propagating modes $u(z)$ and $v(z)$, respectively, 
\begin{equation}\label{efield_ansatz}
  E_{x}(z) = u(z)e^{i(\beta z - \omega t)} + v(z)e^{-i(\beta z + \omega t)} + \text{c.c.}
\end{equation}
where $\beta$ is the propagation constant and $\omega$ is the angular frequency. Substituting the ansatz \eqref{efield_ansatz} into the scalar wave equation for the electric field and making a slowly varying envelope approximation and neglecting small and fast oscillating terms \cite{Yariv1973} leads to the coupled mode equations
\begin{align}
    \frac{\partial u}{\partial z} &= \kappa(z)ve^{-2i\Delta_{B} z}\label{cmt1_steady}\\
    \frac{\partial v}{\partial z} &= \kappa(z)ue^{2i\Delta_{B} z}\label{cmt2_steady}.
\end{align}
where the detuning is given by $\Delta_{B} = \beta - \pi/\Lambda_{B}$ and the position dependent coupling strength by $\kappa(z) = \delta n(z)\beta a(z)/(2\bar{n})$. For both Moir\'e grating types the coupling strength goes to zero periodically with the Moir\'e period. 

The group velocity in a lossless medium is given by $v_{g}=\langle S\rangle/\langle U\rangle$  \cite{Chen2010} where $\langle S\rangle$ is the position-averaged Poynting vector and $\langle U\rangle$ is the position-averaged energy density. By using the ansatz \eqref{efield_ansatz} and denoting the phase velocity as $v_{p}$, it can be shown that the group velocity can be written as \cite{PSBG}
\begin{equation}\label{gv}
    v_{g} = v_{p}\frac{\int_0^{L} dz \; \lvert u\rvert^{2} - \lvert v\rvert^{2}}{\int_0^{L} dz \; \lvert u\rvert^{2} + \lvert v\rvert^{2}}.
\end{equation}


\section{\label{sec:level3}transmission band}

In the approximation where a Moir\'e grating is modeled by two distinct Bragg gratings, the band edges are given by $\lambda_{1}=\Lambda_{1}(2\bar{n}+\delta n/2)$ and $\lambda_{2}=\Lambda_{2}(2\bar{n}-\delta n/2)$ which gives a transmission bandwidth of
\begin{equation}\label{bragg_tbw}
    \Delta f = \frac{c}{\bar{n}}\bigg(\frac{1}{\Lambda_{s}} - \frac{\delta n}{\lambda_{B}}\bigg).
\end{equation}
The transmission band decreases as either the Moir\'e period or the grating strength is increased. The bandwidth for a Bragg grating bandgap is given by 
\begin{equation}\label{bragg_rbw}
    \Delta f_{B} = v_{p}\delta n/\lambda_{B}.
\end{equation}
The Moir\'e transmission bandwidth will be approximately equal to the Bragg bandgap when the Moir\'e period is equal to 
\begin{equation}\label{moire_period}
    \Lambda_{S} = 2\lambda_{B}/3\delta n.
\end{equation}
For Moir\'e periods less than this, Eq.~\eqref{bragg_tbw} provides a good approximation for the transmission bandwidth. 

\begin{figure}
	\centering
	\begin{subfigure}{0.4\textwidth} 
		\includegraphics[width=\textwidth]{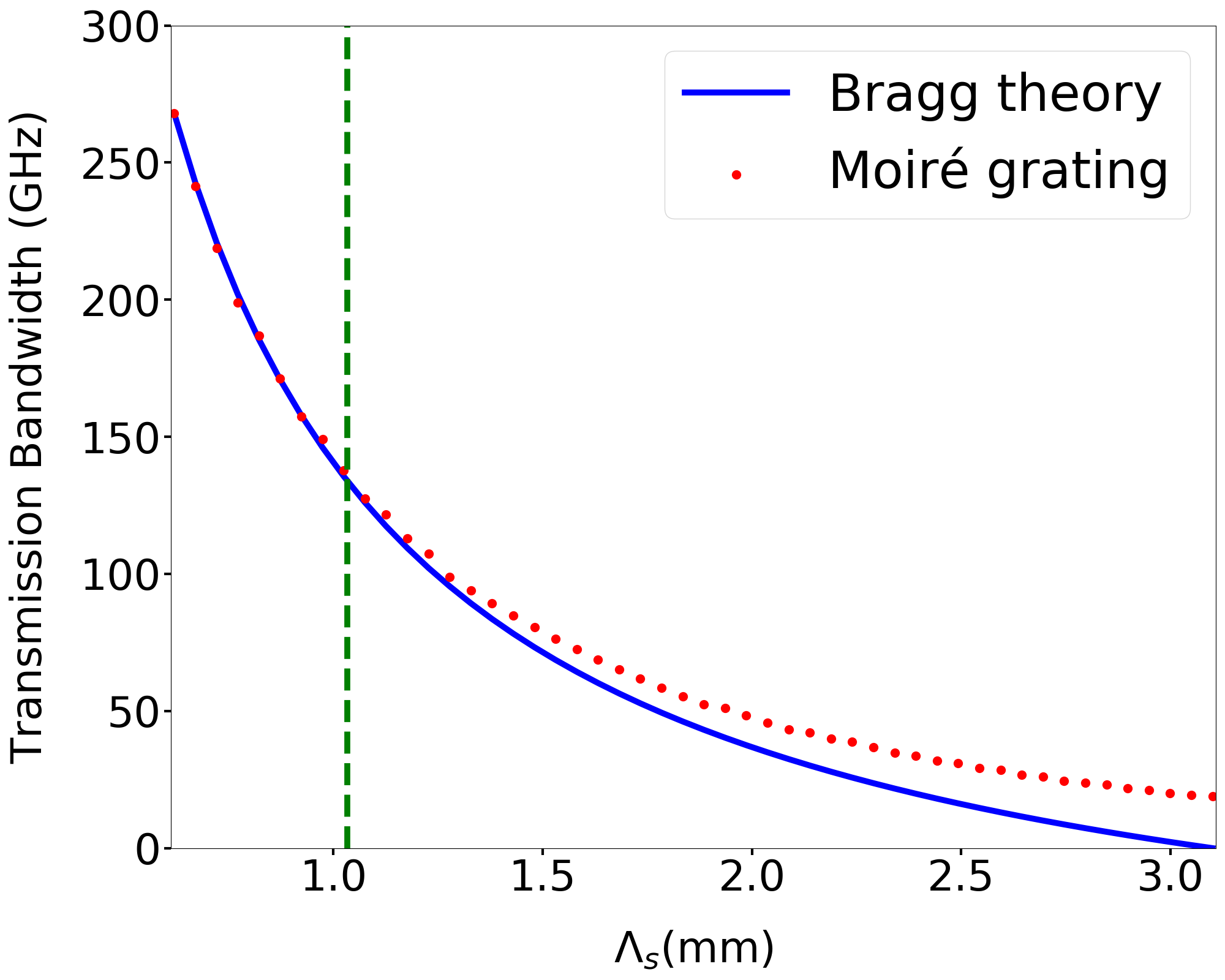}
		\caption{} 
		\label{fig:dsv_tfbw_bragg}
	\end{subfigure}
	\vspace{1em} 
	\begin{subfigure}{0.4\textwidth} 
		\includegraphics[width=\textwidth]{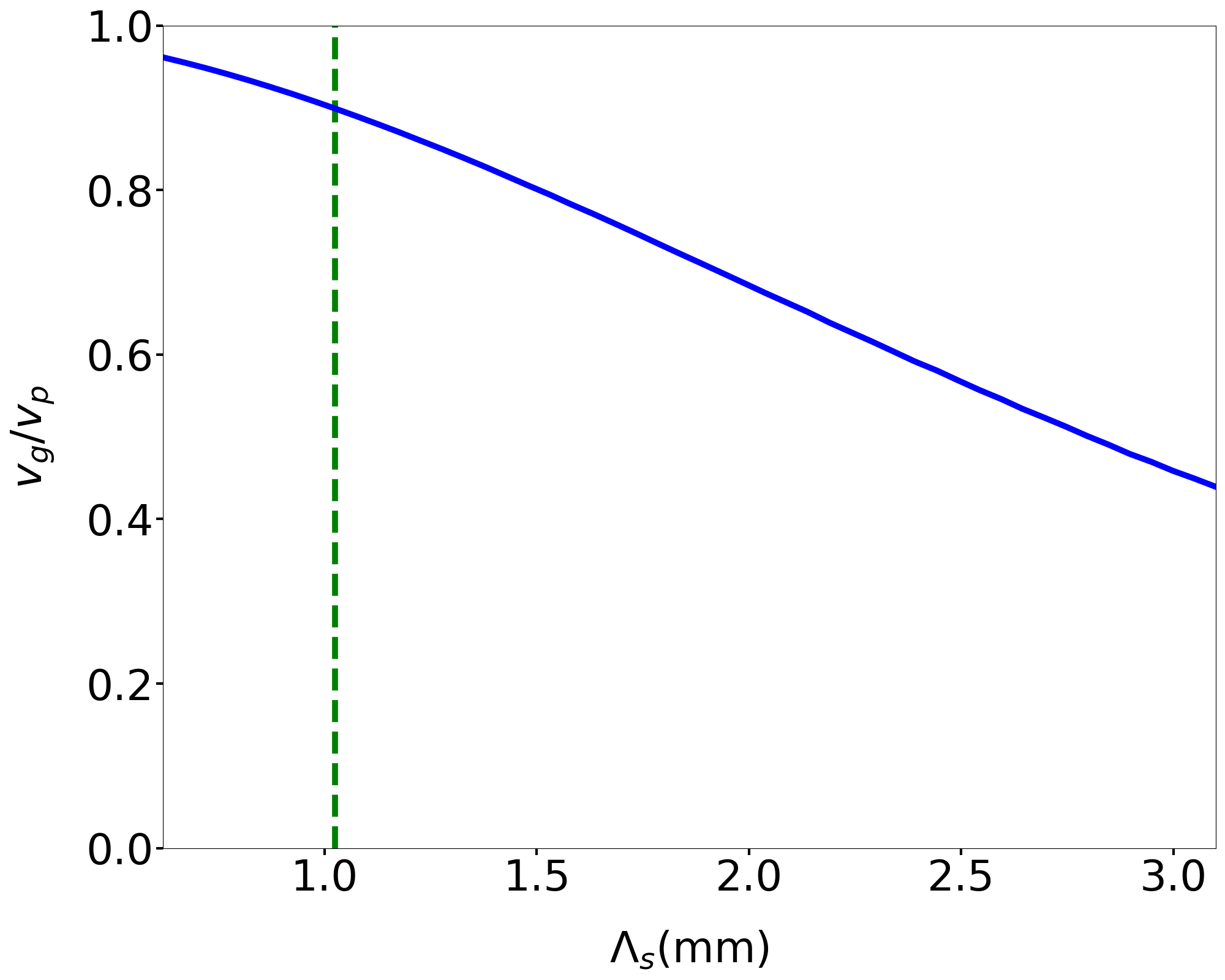}
		\caption{} 
        \label{fig:tfbw_gv}		
	\end{subfigure}
    \caption{(a) Moir\'e transmission bandwidth calculated from Bragg theory and a numerical calculation, and (b) group velocity of a Moir\'e grating versus Moir\'e period $\Lambda_S$. The grating has a Gaussian apodization with parameters $L=20$\, cm, $\lambda_{B} = 1.55\mu$m, $\bar{n} = 1.445$, $\delta n = 10^{-3}$. The vertical green line indicates the Moir\'e period beyond which the approximation of the Moir\'e grating by two Bragg gratings begins to break down.}	
\end{figure}

The actual transmission bandwidth can only be calculated numerically. One way of doing this is to use the position-averaged energy density
\begin{equation}\label{edensity_avg}
    \langle U\rangle = \frac{2\epsilon_{0}\bar{n}^{2}}{L} \int_0^{L} dz \; \lvert u\rvert^{2} + \lvert v\rvert^{2}.
\end{equation}
From Fig.~\ref{fig:gv} it can be seen that the group velocity tends to zero as the wavelength approaches the bandgaps. At these wavelengths the wave is strongly coupled to the medium which leads to an increase in energy density. Therefore the band edges of the transmission band correspond to maxima in $\langle U\rangle$. Figure \ref{fig:dsv_tfbw_bragg} shows a plot of the transmission bandwidths given by Eq.\ \eqref{bragg_tbw} and calculated numerically using the position-averaged energy density from Eq.\ \eqref{edensity_avg}, where the fields $u$ and $v$ were obtained by solving the coupled mode equations \eqref{cmt1_steady} and \eqref{cmt2_steady} using a 4th-5th order Runge-Kutta method with initial conditions $u(0)=1$ and $v(L)=0$.
The vertical line in  Fig.~\ref{fig:dsv_tfbw_bragg} corresponds to the Moir\'e period which gives a transmission bandwidth equal to the Bragg bandgap given by Eq.\ \eqref{moire_period}. The transmission bandwidth becomes a slow light resonance for Moir\'e periods greater than this as shown in Fig.~\ref{fig:tfbw_gv}.

\section{\label{sec:level4}Time varying coupling strength}

We now consider a situation in which $\delta n$ is increased in time in a Moir\'e-type grating while a pulse is propagating through the transmission band. The bandgap bandwidth of a Bragg grating is given by Eq.\ \eqref{bragg_rbw} and is proportional to the grating strength so that as $\delta n$ increases the bandgap increases. Thus, in a Moir\'e grating the transmission bandwidth decreases due to the broadening of the dual bandgaps with increasing grating strength. From Figs.\ \ref{fig:dsv_tfbw_bragg} and \ref{fig:tfbw_gv} we can see that a smaller transmission bandwidth results in a lower group velocity. Therefore increasing $\delta n$ in a Moir\'e type grating produces a slower group velocity within the transmission band. If a pulse is localized within the transmission band when $\delta n$ is increased, the pulse will experience a slower group velocity than the initial group velocity of the grating. By continuing to increase $\delta n$ the pulse can be brought to a virtual standstill with an ultra low group velocity known as ``stopped light''.

This behavior can again be modeled using coupled mode theory but with two modifications. Firstly the electric field ansatz has to allow for forward and backward modes that are also time-dependent,
\begin{equation}\label{efield_ansat_td}
  E_{x}(t,z) =  u(t,z)e^{i(\beta z - \omega t)} + v(t,z)e^{-i(\beta z + \omega t)} + c.c,
\end{equation}
and secondly the refractive index also has to be modified so that the grating strength becomes both time and space dependent $\delta n(t,z)$, %
\begin{equation}\label{moire_linear_td}
    n(t,z) = \bar{n} + \delta n(t,z) a(z)\sin(\alpha_{B}z).
\end{equation}
Note that for simplicity we will ignore the wavelength-dependence of the refractive index in the following, i.e., we neglect the material group velocity dispersion, since it is negligible compared to the grating-induced dispersion for the bandwidths of interest.
Substituting the new expressions \eqref{efield_ansat_td} and \eqref{moire_linear_td} into the wave equation and making rotating wave, slowly varying and neglecting small term approximations \cite{Yariv1973} gives the coupled mode equations
\begin{align}
    \frac{1}{\bar{v}_{p}}\frac{\partial u}{\partial t} + \frac{\partial u}{\partial z}  &= \kappa(t,z) ve^{-2i\Delta_{B} z}\label{cmt1_td}\\
    \frac{1}{\bar{v}_{p}}\frac{\partial v}{\partial t} - \frac{\partial v}{\partial z}  &= -\kappa(t,z) ue^{2i\Delta_{B} z}\label{cmt2_td}
\end{align}
where the coupling strength is given by $\kappa(t,z) = \delta n(t,z)\beta a(z)/(2\bar{n}$). These equations are the time-dependent generalizations of the steady state coupled modes equations \eqref{cmt1_steady} and \eqref{cmt2_steady}. 

We write the time-dependent grating strength as
\begin{equation}\label{dn_td}
    \delta n(t,z) = \delta nf(z)\big(1 + \mu g(t)\big),
\end{equation}
where $g(t)$ is a temporal windowing function which smoothly varies between 0 and 1 and where $\mu$ is a parameter that sets the magnitude of the variation and is constrained such that $\mu\geq0$. When $\mu$ is set to zero we recover the grating strength \eqref{grating_strength}. If it is greater than zero the time variation is controlled by the temporal envelope. When the envelope $g(t)$ reaches its maximum value of 1, the grating strength will be double its original value when $\mu=1$, three times its original value at $\mu=2$ and so on.

For a pulse propagating through the transmission band of a Moir\'e-type grating, described by Eqs.\ \eqref{cmt1_td} and \eqref{cmt2_td}, the temporal variation given by Eq.\ \eqref{dn_td} will induce a group velocity change. Therefore we want any temporal window to be uniformly applied along the spatial length of a pulse to avoid any pulse distortion due to different portions of the pulse having different group velocities. As previously discussed, the grating must have an apodization $f(z)$ in order to suppress reflections within the transmission band. Consequently, the apodization needs to have a flat-top profile. A temporal variation of the grating strength can then be applied when a pulse is localized within the flat-top portion of the grating where the grating strength is spatially uniform. 

\begin{figure}[!ht]
	\centering
	\begin{subfigure}{0.45\textwidth}
		\includegraphics[width=\textwidth]{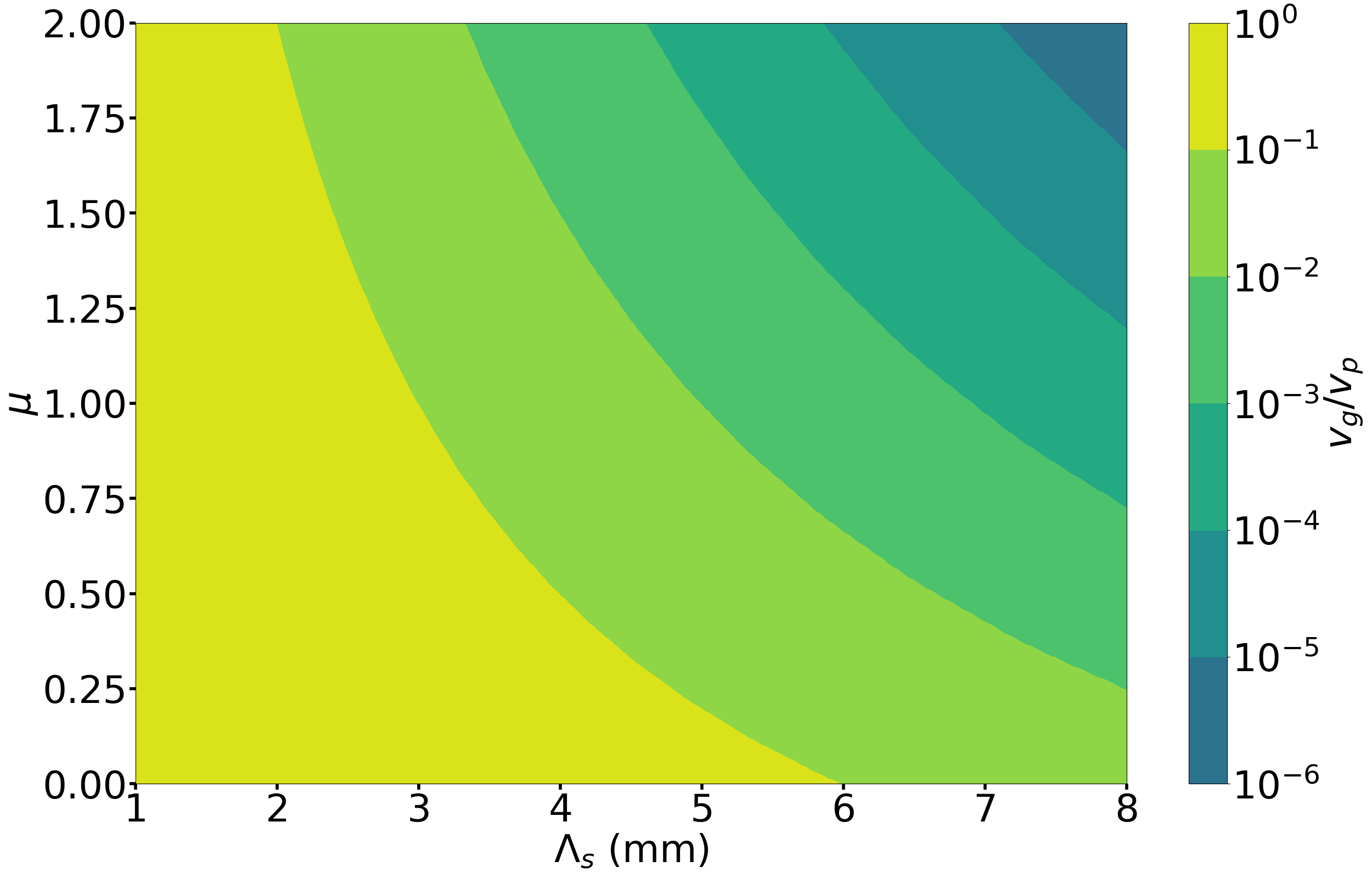}
        \caption{\label{fig:dsv_mu_gv}}
	\end{subfigure}
	\begin{subfigure}{0.45\textwidth}
		\includegraphics[width=\textwidth]{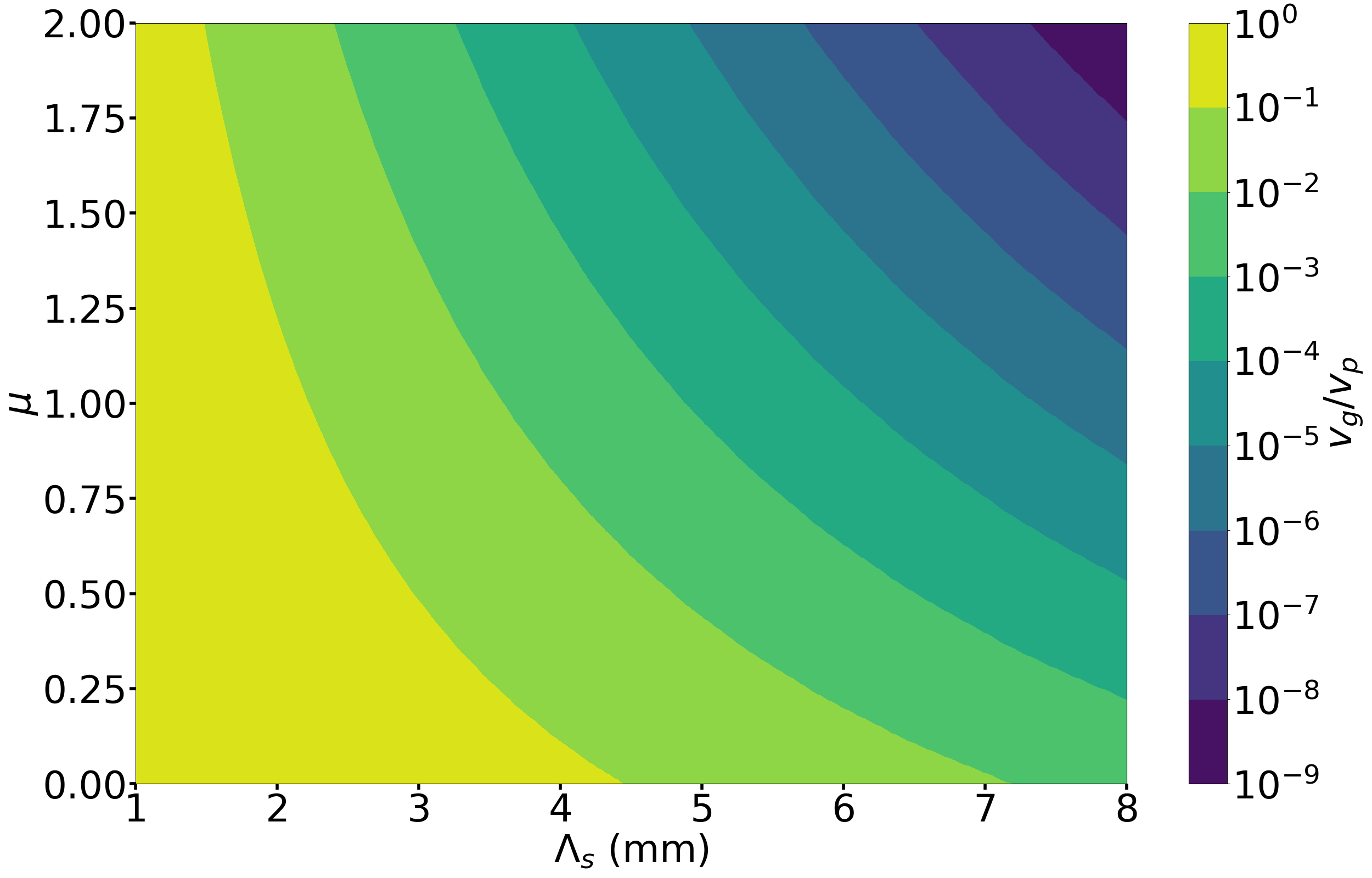}
        \caption{\label{fig:dsv_mu_gv_pi}}
	\end{subfigure}
    \caption{Group velocity of (a) a Moir\'e grating and (b) a $\pi$ phase-shifted grating versus $\Lambda_{s}$ and $\mu$. The grating has a raised cosine apodization with 70\% flat top with parameters $L=20$cm, $\lambda_{B} = 1.55\mu$m, $\bar{n} = 1.445$, $\delta n = 10^{-3}$.}
\end{figure}

Figures \ref{fig:dsv_mu_gv} and \ref{fig:dsv_mu_gv_pi} show how the group velocity varies with the Moir\'e period $\Lambda_S$ and with the parameter $\mu$ for a Moir\'e grating and a $\pi$ phase-shifted grating respectively. As $\Lambda_{s}$ and $\mu$ increase the group velocity decreases. The $\pi$ phase-shifted grating has a higher average grating strength than the Moir\'e grating which means it can achieve a lower group velocity for the same parameters as the Moir\'e grating as shown in the figures. However, in both cases we can expect orders of magnitude reductions in the group velocity when the grating strength is changed by relatively small factors $\mu$. 

To simulate pulse propagation through the transmission band of a dynamic Moir\'e grating we substitute the expression \eqref{dn_td} for the time-dependent grating strength into the coupled mode equations \eqref{cmt1_td} and \eqref{cmt2_td} and work with a carrier wavelength equal to the Bragg wavelength so that the detuning $\Delta_{B}$ is zero. 

\begin{figure}[!ht]
	\centering
	\begin{subfigure}{0.46\textwidth} 
		\includegraphics[width=\textwidth]{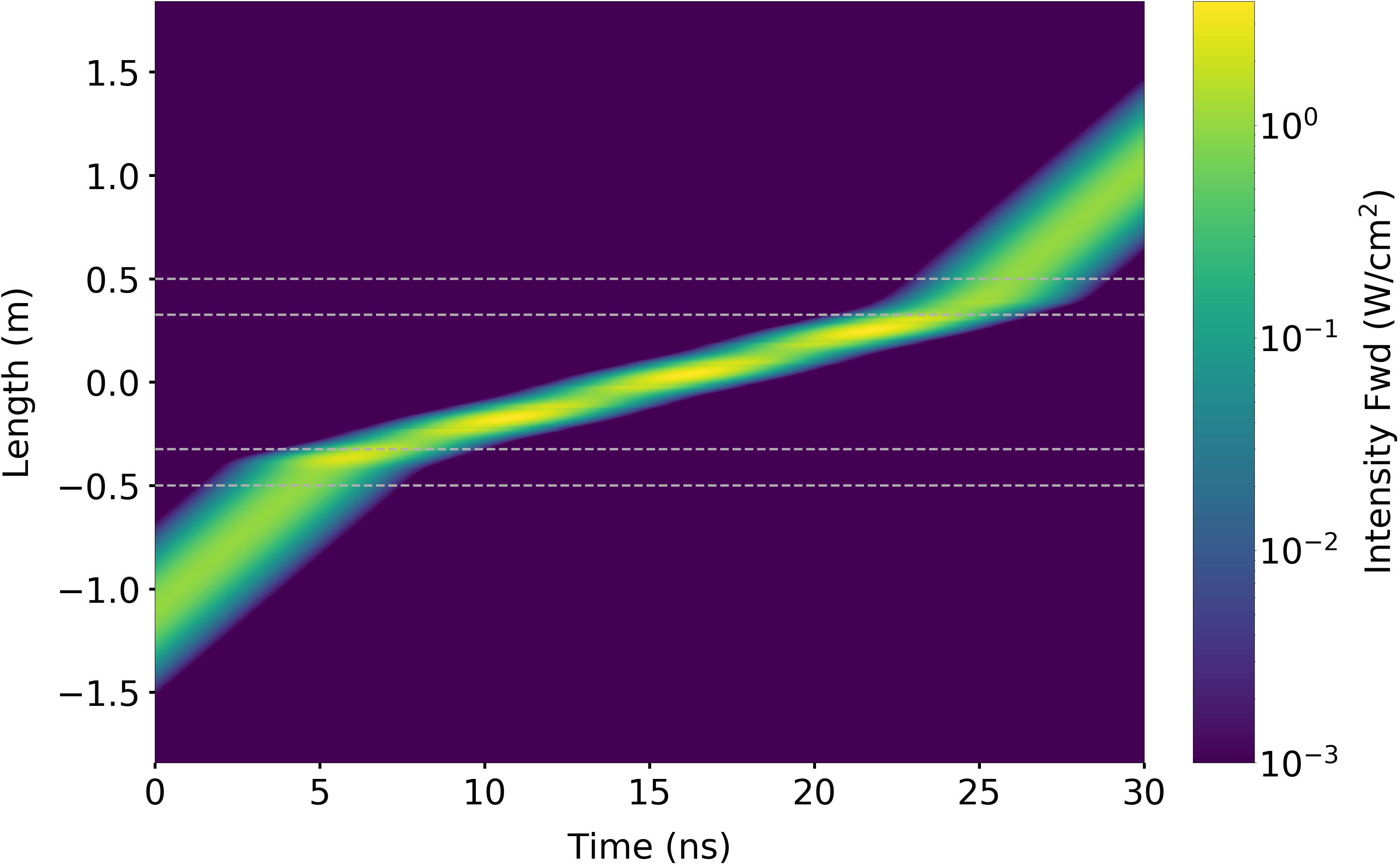}
		\caption{} 
        \label{fig:intensity_fwd_log}
	\end{subfigure}
	\vspace{1em} 
	\begin{subfigure}{0.46\textwidth} 
		\includegraphics[width=\textwidth]{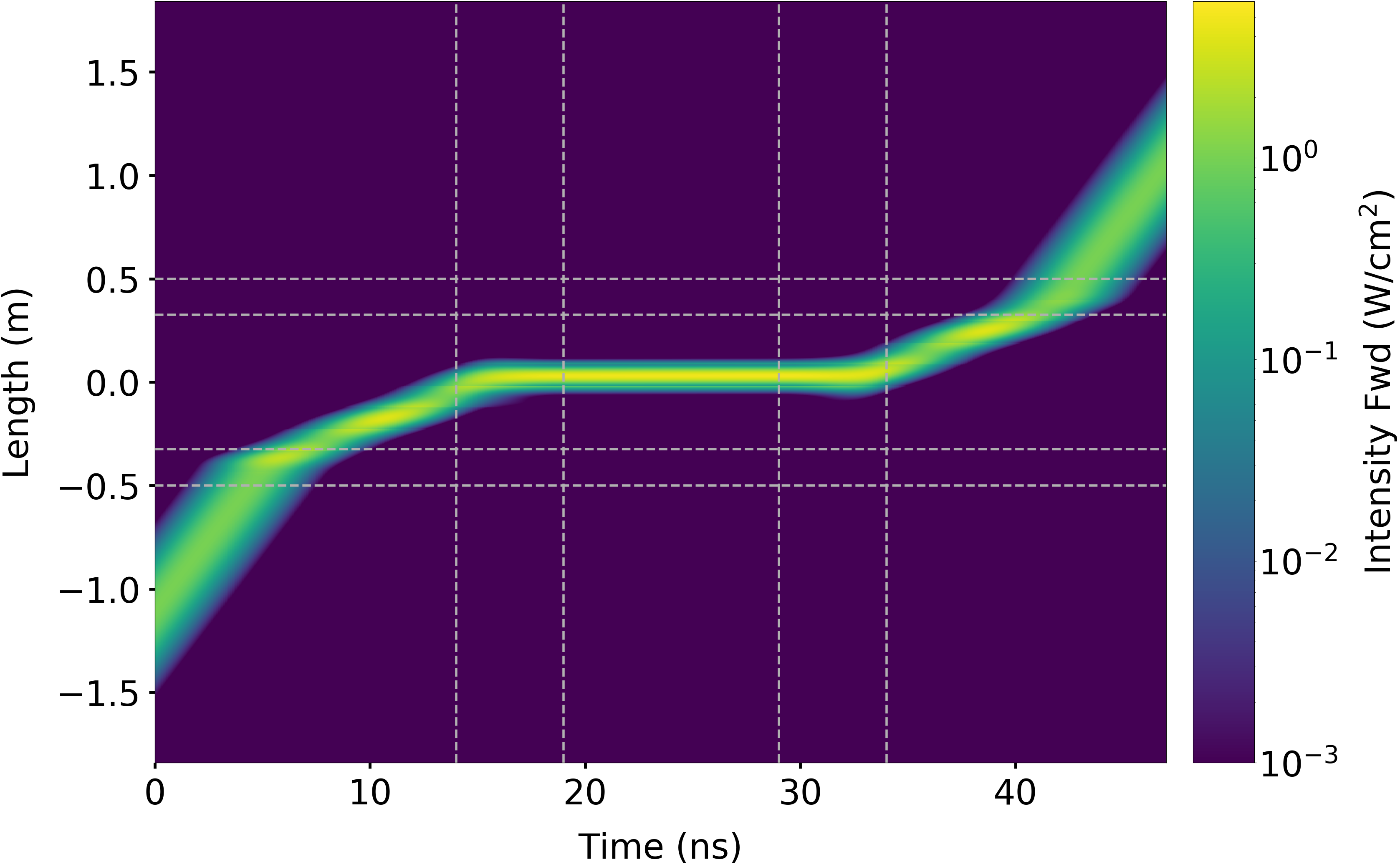}
		\caption{} 
        \label{fig:intensity_fwd_log_stop}
	\end{subfigure}
    \caption{Intensity distribution is space and time of a pulse propagating through (a) a static Moir\'e grating and (b) a dynamic Moir\'e grating with $\mu = 2$. The grating has a raised cosine apodization with 65\% flat top with parameters $L=1$\,m, $\Lambda_{S} = 4$mm, $\lambda_{B} = 1.55\mu$m, $\bar{n} = 1.445$, $\delta n = 10^{-3}$. The pulse has FWHM spectral bandwidth of 320MHz with carrier wavelength $\lambda_{B}$.}
    \label{fig:pulse_stop}
\end{figure}

We performed the simulation by using a Gaussian pulse with a full width at half maximum (FWHM) bandwidth of 320MHz and a 1 meter long grating with a raised cosine apodization profile with a 70\% flat top. We also used a raised cosine for the temporal window function $g(t)$ with a switching time $\Delta\tau = 5\mu$s which is the time taken for the function to switch between 0 and 1. The simulation was then performed by integrating Eqs.\ \eqref{cmt1_td} and \eqref{cmt2_td} in time using a 4th-5th order Runge-Kutta method with the spatial derivatives calculated using spectral methods. Figures \ref{fig:intensity_fwd_log} and \ref{fig:intensity_fwd_log_stop} show the simulated pulse intensity in space and in time with $\mu=0$ and $\mu=2$, respectively. With $\mu=0$ there is no time variation and the pulse propagates through a static Moir\'e grating. The horizontal lines indicate the start and end of the grating and the region where the apodization profile has a constant value of 1. Inside the grating the group velocity is reduced by approximately one order of magnitude. In Fig.\ \ref{fig:intensity_fwd_log_stop} the time variation has magnitude $\mu=2$ and the temporal window switches on when the pulse is localized in the constant region of the apodization. The vertical lines indicate the start and end of the windowing function and the region where the windowing has a constant value of 1. Switching on the windowing causes a further reduction in the group velocity so that the overall group velocity is reduced by approximately three orders of magnitude as seen in Fig.\ \ref{fig:dsv_mu_gv}. When the window is switched off the group velocity returns back to that of the static Moir\'e grating and when the pulse exits the grating it returns to its original group velocity so that the whole process happens adiabatically. 

Consequently, we find that dynamically increasing the grating coupling strength with a pulse localized inside the grating decreases the group velocity of the pulse. The size of the decrease is dependent on the change in grating modulation $\delta n$. For a fixed Moir\'e period, larger values of $\delta n$ result in a narrower transmission bandwidth which in turn means a smaller pulse acceptance bandwidth. In a dynamic Moir\'e grating, on the other hand, the bandwidth is determined by the initial, weak grating while the slowdown is determined by the strong grating after switching the grating strength, thereby beating the delay-bandwidth product limit of a static grating. This is made possible by the adiabatic compression of the pulse spectrum in the time-dependent grating as discussed in the following section.


\section{\label{sec:level5}Bandwidth modulation}

When a pulse with a spectral bandwidth within the transmission band enters the Moir\'e grating, its leading edge will experience a smaller group velocity and travel slower than the trailing edge of the pulse outside the grating. If a pulse has a FWHM spatial length $\Delta z$, then $\Delta z$ will be compressed by a factor of $v_{g}/v_{0}$ as the pulse enters the grating, where $v_{0}$ is the group velocity outside the grating and $v_{g}$ is the group velocity inside the grating. By the same argument in reverse, as a pulse exits the grating $\Delta z$ is broadened by a factor $v_{0}/v_{g}$. The net result is that the two factors cancel and the pulse emerges from the grating unaltered. The compression and decompression is caused by the grating dispersion and as such the pulse spectrum is unaltered.

If the group velocity is varied in time to $\tilde{v}_{g}$ by switching on a windowing function when the pulse is localized inside the grating, then $\Delta z$ is not compressed further as the variation happens across the whole pulse at the same time. If we now consider the pulse exiting the grating without the windowing function being switched off, the change in group velocity inside and outside the grating is different from when the pulse entered the grating and consequently $\Delta z$ is scaled by a factor $v_{g}/\tilde{v}_{g}$. Through a Fourier transform the FWHM bandwidth $\Delta f$ of the pulse is related to $\Delta z$ by $\Delta f = 4c\ln(2)/(\pi\Delta z$). Subsequently the pulse spectrum for the pulse exiting the grating is scaled by a factor $\tilde{v}_{g}/v_{g}$ so that
\begin{equation}\label{spectrum_change}
    \Delta \tilde{f} = \frac{\tilde{v}_{g}}{v_{g}}\Delta f
\end{equation}
where $\Delta \tilde{f}$ is the FWHM bandwidth of the pulse exiting the grating.

\begin{figure}[!ht]
	\centering
	\begin{subfigure}{0.45\textwidth}
		\includegraphics[width=\textwidth]{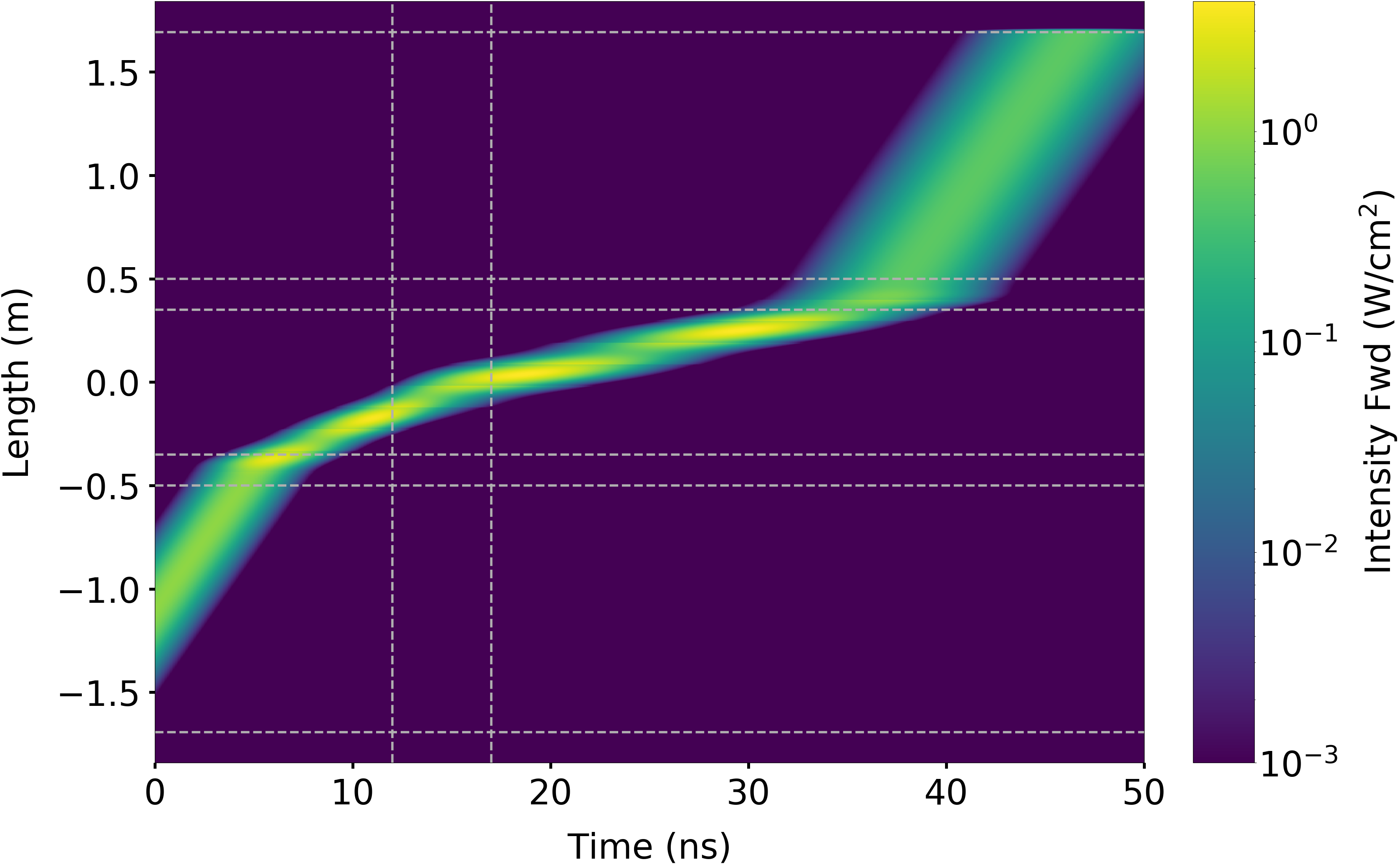}
		\caption{\label{fig:bw_compress_pulse}}
	\end{subfigure}
	\begin{subfigure}{0.45\textwidth}
		\includegraphics[width=\textwidth]{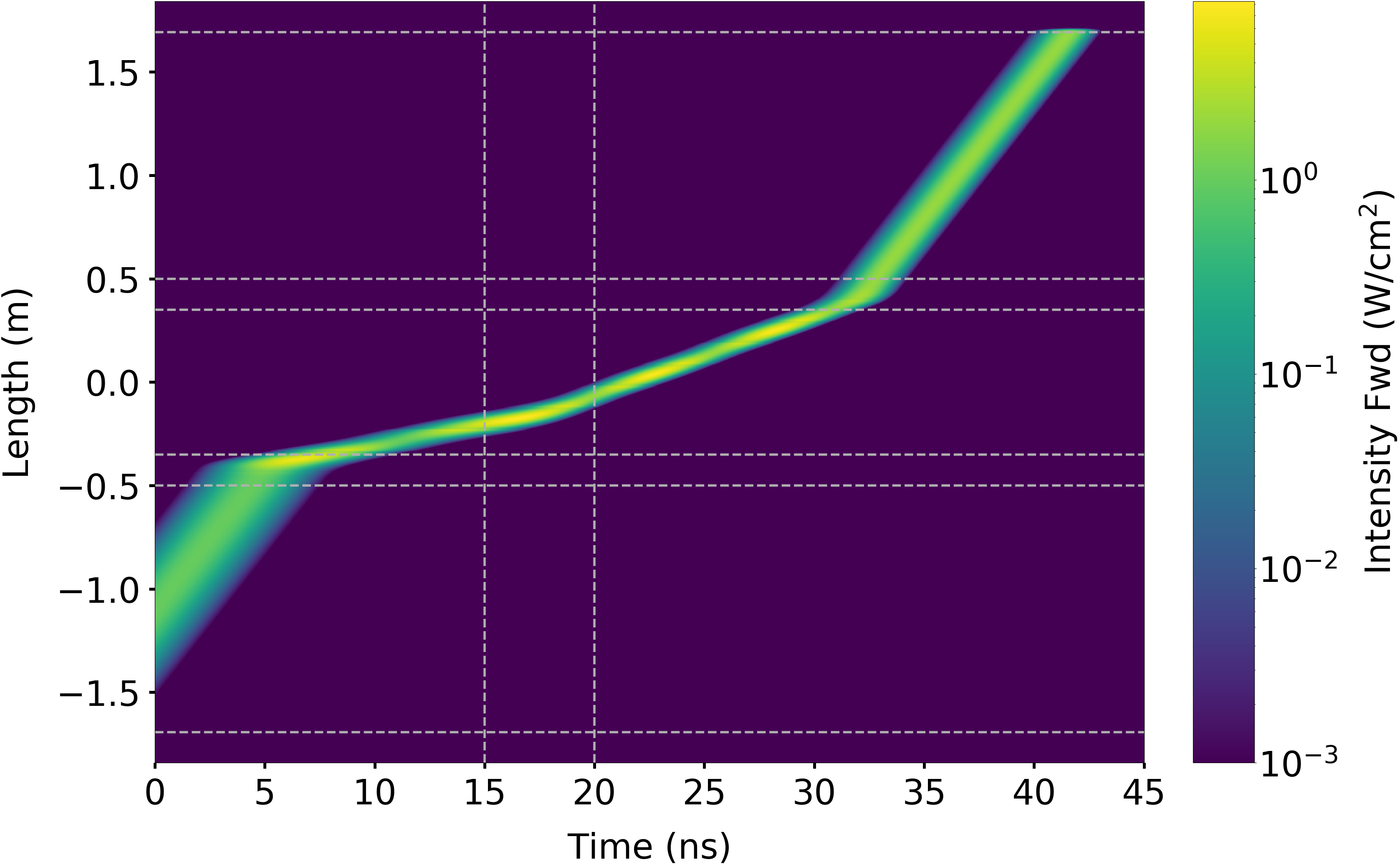}
		\caption{\label{fig:bw_stretch_pulse}}
	\end{subfigure}
	\begin{subfigure}{0.4\textwidth}
		\includegraphics[width=\textwidth]{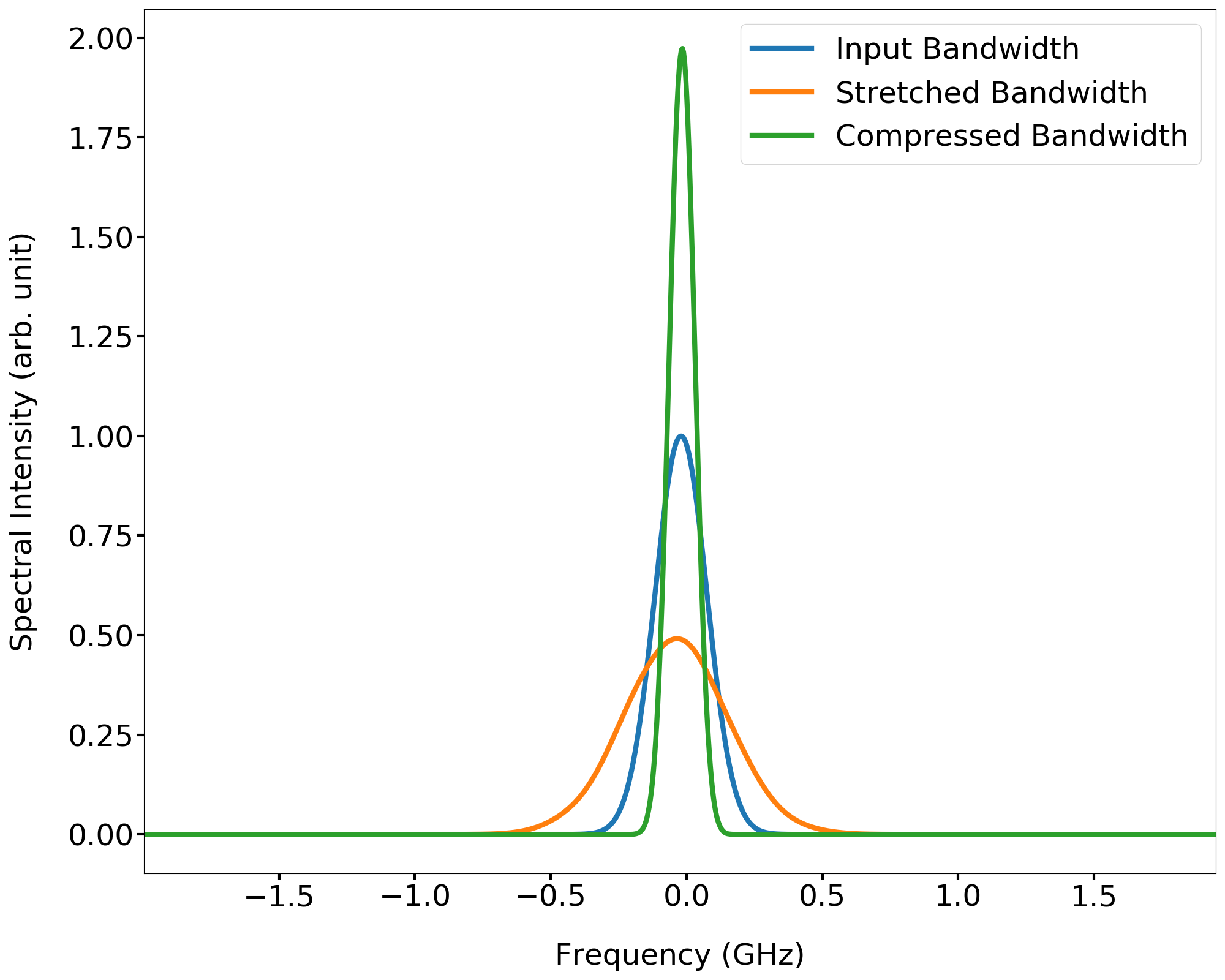}
		\caption{} 
        \label{fig:spectral_intensity_compare}		
	\end{subfigure}	
    \caption{Intensity profile of a pulse propagating through (a) a dynamic spectrum-compressing Moir\'e grating and (b) a dynamic spectrum-broadening Moir\'e grating. (c) Comparison of the input and output pulse spectra from (a) and (b). Parameters as in Fig.\ \ref{fig:pulse_stop}. The time variation magnitude has value $\mu=0.33$ in both (a) and (b).}
\end{figure}

We can thus consider a situation in which the group velocity is not varied symmetrically but either increased or decreased without being subsequently changed back to its original value. If the group velocity is altered such that it is larger than the original group velocity, then the spectrum will be broadened. Correspondingly, if the altered group velocity is smaller than the original group velocity then the spectrum will be compressed.

We simulated this behavior by again integrating equations \eqref{cmt1_td} and \eqref{cmt2_td} using the same methods as previously described but using windowing functions which either only switch on or switch off. Figure \ref{fig:bw_compress_pulse} shows a pulse propagation simulation where the windowing function is initially switched off and then switches on when the pulse is localized in the grating, which causes the pulse spectrum to compress. The windowing remains on as the pulse exits the grating and the compressed spectrum causes the pulse the spatially broaden as it leaves the grating. In contrast, figure \ref{fig:bw_stretch_pulse} has a windowing function that is initially switched on and subsequently switched off when the pulse is localized inside the grating. This causes the pulse spectrum to broaden such that when the pulse exits the grating it is spatially compressed.  

\begin{figure}[!ht]
    \includegraphics[scale=.12]{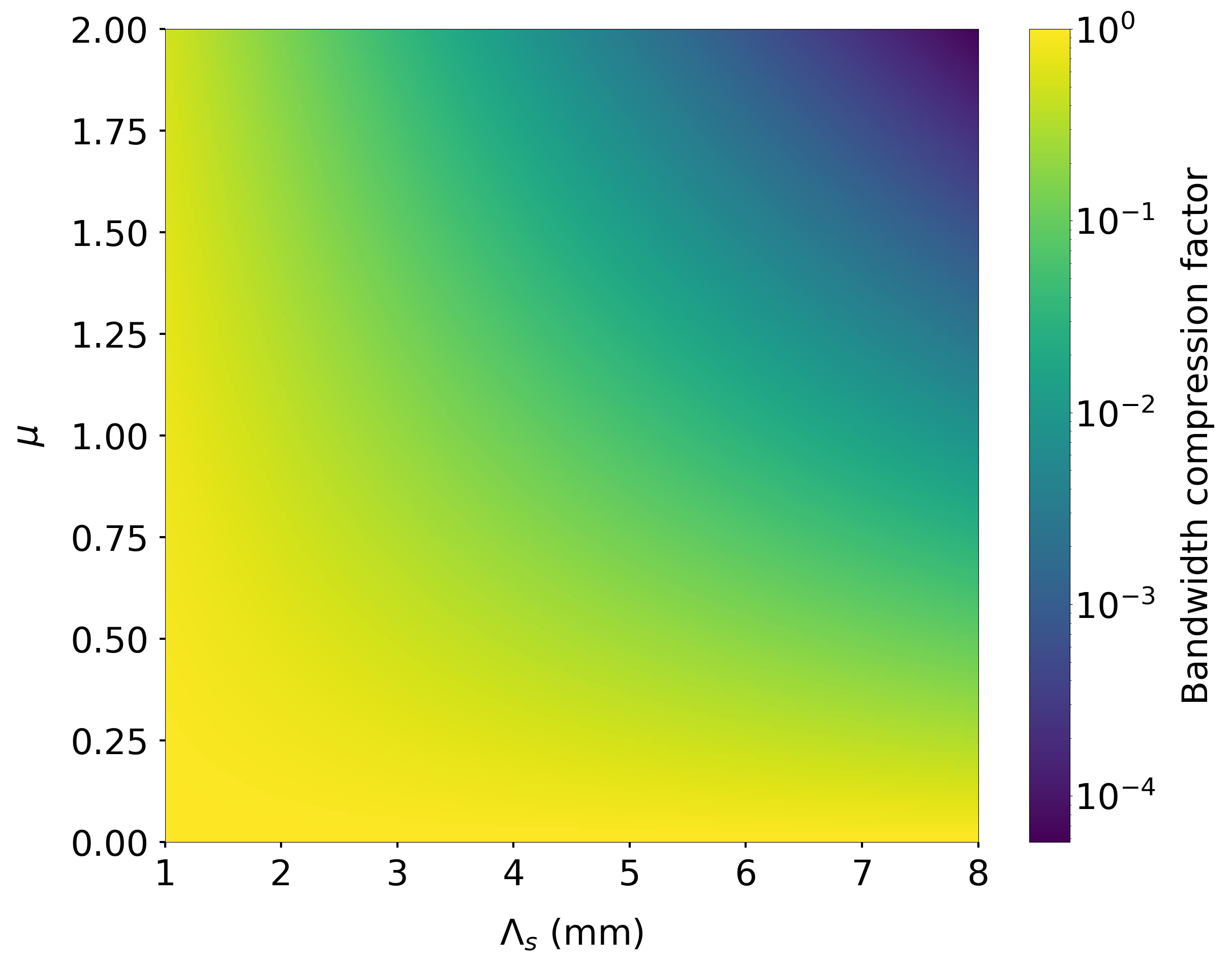}
    \centering
    \caption{Bandwidth compression factor depending on $\Lambda_{s}$ and $\mu$ in a dynamic Moir\'e grating with parameters $L=20$cm, $\lambda_{B} = 1.55\mu$m, $\bar{n} = 1.445$, $\delta n = 10^{-3}$.}
    \label{fig:bw_compress_scan}
\end{figure}

In both simulations the time variation has magnitude $\mu=0.33$ which causes a bandwidth compression of one half in Fig.\ \ref{fig:bw_compress_pulse} and a spectrum broadening of a factor two in Fig.\ \ref{fig:bw_stretch_pulse}. Figure \ref{fig:spectral_intensity_compare} shows a comparison of the pulse spectrum upon leaving the grating in cases when the grating is static (equivalent to the input pulse spectrum), when the spectrum is compressed and when the spectrum is broadened. Figure \ref{fig:bw_compress_scan} shows how the spectrum compression factor varies with $\Lambda_{S}$ and $\mu$ in the case of a windowing function that is switching on during pulse propagation.  The broadening from switching the windowing off is the reciprocal of the spectrum compression factor. By comparing Fig.\ \ref{fig:bw_compress_scan} with Figs.\ \ref{fig:dsv_mu_gv} and \ref{fig:dsv_mu_gv_pi}, we note that ultra-slow group velocities correspond to extreme spectrum compression. 

It is therefore possible to stretch or compress the pulse by several orders of magnitude by switching the grating strength dynamically by factors of two or three. Moreover, as can be seen in Fig.\ \ref{fig:spectral_intensity_compare}, the spectrum maintains its Gaussian shape because the refractive index is changed uniformly across the entire pulse in space. This is in sharp contrast to traditional spectral broadening methods based, e.g., on self-phase modulation in a Kerr medium \cite{Agrawal1989}.

\section{\label{sec:level6}Realization of Dynamic Moir\'e Gratings}

Dynamic gratings have been proposed by using various approaches such as an erbium-doped fiber \cite{Frisken1992}, Brillouin scattering \cite{Song2008} and excited free-carriers \cite{Sivan2015}. The approach we propose here is to use an electro-optic grating. Such a grating is created by applying an external quasi-static electric field to a periodically poled $\chi^{(2)}$ medium \cite{730484}. Through the interaction with the external field the poling induces a Bragg grating in the linear refractive index as will be shown in the following. 

\begin{figure}[!ht]
    \includegraphics[scale=.3]{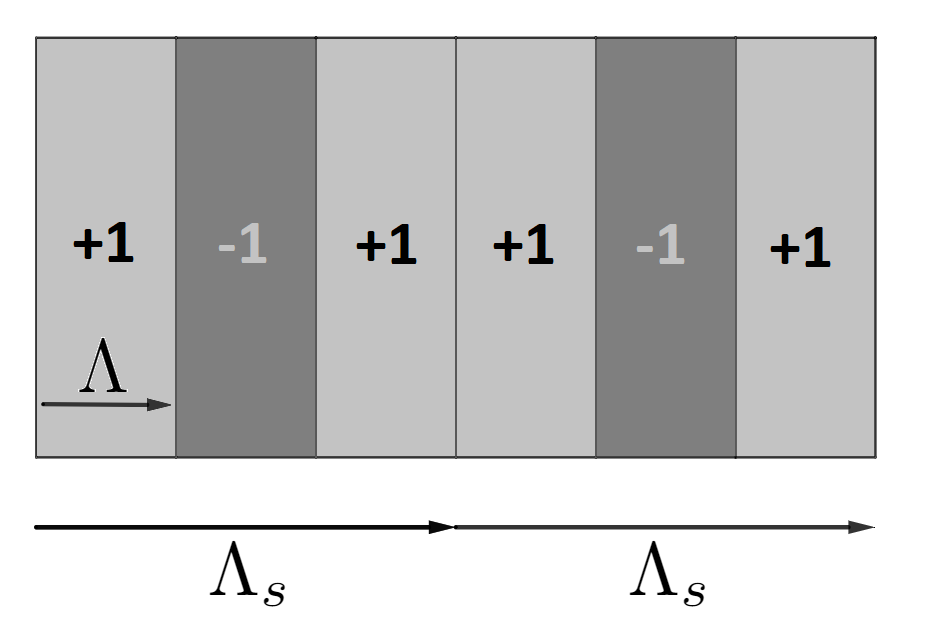}
    \centering
    \caption{An example $\chi^{(2)}$ Moir\'e grating with a Moir\'e period that is three times the length of the poling period. The effect of the Moir\'e period is to create a single double-length poling period after each Moir\'e period.}
    \label{fig:chi2_grating_example}
\end{figure}

The $\chi^{(2)}$ poling process only allows the sign of the $\chi^{(2)}$ nonlinearity to be changed. Therefore a $\chi^{(2)}$ poling profile for an electro-optic Moir\'e grating is given by
\begin{equation}\label{chi2_polling_sgn}
    \chi^{(2)}(z) = \chi^{(2)}\text{sgn}\big[\sin(\alpha_{S}z)\big]\text{sgn}\big[\sin(\alpha_{B}z)\big]
\end{equation}
The Moir\'e and Bragg terms $\sin(\alpha_{S}z)$ and $\sin(\alpha_{B}z)$ respectively are determined by the poling where $\alpha_{B}=2\pi/\Lambda$ and $\Lambda$ is the poling period. 
If the Moir\'e period is an integer multiple of the poling period, the term $\text{sgn}\big[\sin(\alpha_{S}z)\big]$ will introduce $\pi$ phase shifts into the $\text{sgn}\big[\sin(\alpha_{B}z)\big]$ term. The $\pi$ phase shifts have the effect of creating a single double-length poling period after each Moir\'e period. An illustration of an example $\chi^{(2)}$ Moir\'e grating profile is given in Fig.\ \ref{fig:chi2_grating_example}. Gratings of this type have previously been studied for multi-wavelength conversion in quasi-phase matched nonlinear crystals \cite{Fradkin-Kashi1999,Chou1999}.

The harmonic content of the square waves in Eq.~\eqref{chi2_polling_sgn} can be understood by expanding them as a Fourier series
\begin{equation}\label{chi2_polling_fourier}
    \chi^{(2)}(z) = \frac{4\chi^{(2)}}{\pi}\text{sgn}\big[\sin(\alpha_{S}z)\big]\sum_{m=1}^{\infty}\frac{\sin\big(\alpha_{B}(2m-1)z\big)}{2m-1}
\end{equation}
The higher order Fourier components have larger denominators which reduce the overall strength of $\chi^{(2)}$, therefore we will limit our analysis to only the first component in the following. The $\chi^{(2)}$ grating profile can thus be written as
\begin{equation}\label{chi2_polling}
    \chi^{(2)}(z) = \frac{4\chi^{(2)}}{\pi}\text{sgn}\big[\sin(\alpha_{S}z)\big]\sin\big(\alpha_{B}z\big).
\end{equation}

A set of coupled mode equations for the grating can be derived by substituting the $\chi^{(2)}$ grating \eqref{chi2_polling} into the nonlinear wave equation
\begin{equation}\label{wave_eq_nonlin}
    \frac{\partial^{2}E}{\partial z^{2}} - \frac{n^{2}}{c^{2}}\frac{\partial^{2}E}{\partial t^{2}} = \mu_{0}\frac{\partial^{2}P^{NL}}{\partial t^{2}}
\end{equation}
with the nonlinear polarization $P^{NL} = \epsilon_{0}\chi^{(2)}(z)E^{2}$. 

The electric field ansatz needs to be modified from Eq.~\eqref{efield_ansat_td} to include an external quasi-static field $E_{D}(t,z)$ to enable time-variation of the grating strength,
\begin{equation}\label{efield_ansat_chi2}
  E(t,z) = E_{D} + ue^{i(\beta z - \omega t)} + ve^{-i(\beta z + \omega t)} + c.c.
\end{equation}
The linear part of Eq.~(\ref{wave_eq_nonlin}) can be decomposed into forward and backward transport equations for the respective modes which gives 
\begin{align}
    \frac{e^{i(\omega t-\beta z)}}{2i\beta}\bigg(\frac{\partial^{2}E}{\partial z^{2}} - \frac{n^{2}}{c^{2}}\frac{\partial^{2}E}{\partial t^{2}}\bigg) &\approx \frac{1}{v_{p}}\frac{\partial u}{\partial t} + \frac{\partial u}{\partial z}\label{linear_decomp1}\\
    \frac{e^{-i(\omega t+\beta z)}}{2i\beta}\bigg(\frac{\partial^{2}E}{\partial z^{2}} - \frac{n^{2}}{c^{2}}\frac{\partial^{2}E}{\partial t^{2}}\bigg) &\approx \frac{1}{v_{p}}\frac{\partial v}{\partial t} - \frac{\partial v}{\partial z}\label{linear_decomp2}.
\end{align}
The decomposition is an approximation that is achieved by neglecting small and fast oscillating frequency terms and taking a slowly varying envelope approximation which are the standard approximations used to derive coupled mode equations \cite{Yariv1973}. Making the same decomposition and approximations to the nonlinear part of the wave equation yields
\begin{align}
  \frac{e^{i(\omega t-\beta z)}}{2i\beta}\mu_{0}\frac{\partial^{2}P^{NL}}{\partial t^{2}} &\approx \kappa^{(2)}\text{sgn}\big[\sin(\alpha_{S}z)\big]ve^{-2i\Delta_{B} z}\label{non_linear_decomp1}\\
  \frac{e^{-i(\omega t+\beta z)}}{2i\beta}\mu_{0}\frac{\partial^{2}P^{NL}}{\partial t^{2}} &\approx -\kappa^{(2)}\text{sgn}\big[\sin(\alpha_{S}z)\big]ue^{2i\Delta_{B} z}\label{non_linear_decomp2}
\end{align}
where the detuning is the same as for the linear grating and the nonlinear coupling is given by
\begin{equation}\label{kappa_2}
  \kappa^{(2)}(t,z) = \frac{2\beta\chi^{(2)}E_{D}(t,z)}{\bar{n}^{2}\pi}.
\end{equation}

By comparing the nonlinear coupling $\kappa^{(2)}$ with the linear coupling $\kappa$, the application of the quasi-static field $E_{D}$ has the affect of inducing a nonlinear grating strength given by
\begin{equation}\label{delta_2}
  \delta^{(2)}(t,z) = \frac{4\chi^{(2)}E_{D}(t,z)}{\bar{n}\pi}.
\end{equation}

Equations \eqref{linear_decomp1}, \eqref{non_linear_decomp1}, \eqref{linear_decomp2} and \eqref{non_linear_decomp2} can be combined to yield the same coupled mode equations \eqref{cmt1_td} and \eqref{cmt2_td} as before but with the coupling given by
\begin{equation}\label{coupling}
    \kappa(t,z) = \kappa^{(2)}(t,z)\text{sgn}\big[\sin(\alpha_{S}z)\big].
\end{equation}

Therefore an electro-optic grating can create a realization of a dynamic Moir\'e grating by inducing the grating coupling by an external quasi-static field $E_{D}$. Varying the externally applied field allows for dynamic control of the group velocity within the transmission band. As the group velocity reduction is determined by the overall coupling strength, a linear grating (\ref{moire_linear}) could also be included so that the coupling becomes  
\begin{equation}\label{couplinglin}
    \kappa(t,z) = \kappa\sin(\alpha_{S} z) + \kappa^{(2)}(t,z)\text{sgn}\big[\sin(\alpha_{S}z)\big].
\end{equation}
where the Bragg period is equal to the poling period so that the overall maximum coupling is given by $\kappa + \kappa^{(2)}$.

A realizable device would need to be based on a thin film lithium niobate waveguide \cite{Mackwitz2016} in order to achieve the necessary electric field density needed to generate a sufficiently strong $\kappa^{(2)}$. Current limits of poling technology allow for periods of around 750 nm \cite{Nagy20,Suzuki2013} which would give a Bragg wavelength of approximately 3 $\mu$m which is within the transparency window for lithium niobate. The reported coercive field strength for lithium niobate is 21 V/$\mu$m \cite{Venkatraman97}. Applying a field strength at half the coercive field over a 3 $\mu$m thin film lithium niobate waveguide would induce a grating modulation strength in the order of $10^{-4}$. This is a factor of 10 smaller than in our numerical examples above and a practical realization would therefore involve a smaller slow light resonance which would in turn require a longer device. 


\section{\label{sec:level7}Conclusion}

In conclusion we have investigated dynamic Moir\'e gratings capable of producing slow and stopped light which provide improved delay-bandwidth performance compared to a static Moir\'e grating. This is achieved by increasing and subsequently decreasing the coupling strength of the grating in time. When a pulse is localized within the grating, increasing the coupling strength compresses the pulse spectrum and reduces its group velocity. This process is adiabatically reversible by decreasing the coupling strength to its original value. This means that lower group velocities can be achieved for a given acceptance bandwidth which overcomes the usual delay-bandwidth constraint. We then showed that if the coupling strength is varied in a non-symmetric manner so that is either increased or decreased, then the pulse spectrum can be compressed or broadened once it exits the grating. We finally investigated a possible realization of a dynamic Moir\'e grating by using an electro-optic grating. This works by applying an external quasi-static electric field to a poled $\chi^{(2)}$ medium which induces a grating in the linear refractive index. Varying the strength of the applied field then varies the coupling strength of the induced grating which makes it a viable method for creating a dynamic Moir\'e grating.


\begin{acknowledgements}
The authors acknowledge funding through the UK National Quantum Technologies Programme (EPSRC grant numbers EP/T001062/1, EP/M024539/1) and an EPSRC DTP PhD studentship. We also acknowledge the use of the IRIDIS High Performance Computing Facility, and associated support services at the University of Southampton, in the completion of this work. All data supporting this study are openly available from the University of Southampton repository at https://doi.org/10.5258/SOTON/D1858.
\end{acknowledgements}


\bibliography{dynmoire}

\end{document}